\documentclass{article}

\usepackage[english]{babel}
\usepackage[letterpaper,top=2cm,bottom=2cm,left=3cm,right=3cm,marginparwidth=1.75cm]{geometry}

\usepackage{amsmath}
\usepackage{amsthm}
\usepackage{amssymb}
\usepackage{bbm}
\usepackage{graphicx}
\usepackage{tikz}
\usetikzlibrary{positioning}
\usepackage[colorlinks=true, allcolors=blue]{hyperref}

\DeclareMathOperator{\LL}{\mathcal{L}}
\DeclareMathOperator{\adj}{adj}
\DeclareMathOperator{\supp}{supp}
\DeclareMathOperator{\den}{\mathrm{den}}
\DeclareMathOperator{\lcm}{\mathrm{lcm}}
\DeclareMathOperator{\spn}{\mathrm{span}}

\newcommand{\CC}{\mathbb{C}}
\newcommand{\RR}{\mathbb{R}}

\newtheorem{theorem}{Theorem}[section]
\newtheorem{corollary}[theorem]{Corollary}
\newtheorem{lemma}[theorem]{Lemma}

\theoremstyle{definition}

\theoremstyle{remark}

\title{Perfect State Transfer on Trees in the Normalized Laplacian Quantum Walk}
\author{Laureline Polli\footnote{Department of Mathematics, University of Florida, Gainesville, Florida, USA}}

\begin{document}
\maketitle

\begin{abstract}
We prove that, in the normalized-Laplacian model, perfect state transfer between two vertices of a tree can only happen if they are at distance at most $2$. We show that a conjecture of Coutinho and Baptista, stating that the only trees admitting perfect state transfer are $K_2$ and $P_3$, is true when the two vertices are at odd distance. We then construct infinite families of trees admitting perfect state transfer, disproving their conjecture in the general case.
\end{abstract}

\section{Introduction}
Quantum walks on graphs can be used to model the evolution of a network of qubits restricted to its one-excitation subspace~\cite{christandlPerfectStateTransfer2004}. In this model, the vertices of the graph are identified with the basis states in which exactly one qubit is excited. If the system can start in one of these basis states and evolve to a state where the excitation is perfectly localized at another vertex, we say that we have perfect state transfer. This phenomenon has applications in quantum communication, for example to establish entanglement~\cite{kayReviewPerfectState2010}.
Because of these applications, but also because of the questions it raises in graph theory, particularly spectral graph theory, the classification of graphs that admit perfect state transfer has been an important question in the study of quantum walks~\cite{godsilWhenCanPerfect2012, coutinhoQuantumWalksNormalized2024, pantangiPerfectStateTransfer2026}.

As trees are minimally connected, they seem like natural candidates to be physically implemented if we want to minimize the number of interactions in the system \cite{kayReviewPerfectState2010,coutinhoQuantumWalksNormalized2024}.
Unfortunately, it has been proved that using the adjacency matrix as the Hamiltonian of the walk, the only trees admitting perfect state transfer are $K_2$ and $P_3$ \cite{coutinhoNoPerfectState2024}.
Similarly, using the combinatorial Laplacian, $K_2$ is the only tree that admits perfect state transfer~\cite{coutinhoNoLaplacianPerfect2015}.

A similar result was therefore naturally conjectured for the normalized Laplacian model by Coutinho and Baptista in \cite{coutinhoQuantumWalksNormalized2024}. They conjectured that apart from the paths on two and three vertices, perfect state transfer with respect to the normalized Laplacian doesn't occur in trees. In the same paper, they provided several results supporting this conjecture. In particular, they obtained a strong spectral restriction in the case where perfect state transfer occurs between vertices at odd distance in a tree. Furthermore, they proved that if perfect state transfer occurs at even distance, then the two vertices must be leaves with a common neighbor. 

In this paper, we obtain further results in both the odd and even distance cases. We first prove their conjecture in the odd-distance case: $K_2$ is the only tree admitting perfect state transfer between vertices at odd distance. We then disprove their conjecture in the even-distance case: there is an infinite family of trees admitting perfect state transfer between two vertices at even distance. 
After that, we construct further examples of this phenomenon, which illustrate the difficulty of giving a complete structural classification.

\section{Preliminaries}
Throughout, all graphs are connected, finite, simple, and unweighted. For a graph $G$, we write $A$ for its adjacency matrix and $D$ its degree matrix, the diagonal matrix whose entry $D_{v,v}$ is the degree of vertex $v$. We also write $e_v$ for the basis vector corresponding to vertex $v$.

The \textit{combinatorial Laplacian} of $G$ is defined as $L=D-A$. If $G$ has at least two vertices, its \textit{normalized Laplacian} is a degree-normalized version of the combinatorial Laplacian,
\[\LL=D^{-1/2}LD^{-1/2}=I-D^{-1/2}AD^{-1/2}.\]
This normalization accounts for variations in vertex degree, and the spectrum of $\LL$ is known to encode important properties of the graph. In particular, for a connected graph, $2$ is an eigenvalue of $\LL$ if and only if $G$ is bipartite.

The \textit{continuous-time quantum walk} on $G$ associated with $\LL$ is given by the family of matrices
\[U(t)=e^{it\LL}, \quad t\in \RR.\]
Throughout this paper, we only consider the normalized Laplacian model, so $U(t)$ always denotes $e^{it\LL}$ and all results apply only to this model.
We say that distinct vertices $a$ and $b$ admit \textit{perfect state transfer} (PST) at time $\tau$ if 
\[U(\tau)e_a=\gamma e_b\] for some $\gamma \in \CC$. For any $t\in \RR$, $U(t)$ is unitary so, if there is PST between vertices $a$ and $b$, then $|\gamma|=1$. 
As a simple example, consider the path with two vertices $K_2$. The normalized Laplacian of this graph is 
\[\LL=\begin{pmatrix}
    1 & -1\\
    -1 & 1
\end{pmatrix}.\]
Thus, 
\[U(t)=e^{it\LL}=e^{it} \begin{pmatrix}
    \cos(t) & -i \sin(t)\\
     -i \sin(t) & \cos(t) 
\end{pmatrix}.\]
As $U(\pi/2)e_1=e_2$, there is PST between the two vertices at time $\pi/2$.

Let $\LL =\sum_r \theta_r E_r $ be the spectral decomposition of $\LL$ where the $\theta_r$ are its distinct eigenvalues and $E_r$ is the orthogonal projector onto the eigenspace corresponding to $\theta_r$.
The \textit{eigenvalue support} of a vector $v$ with respect to $\LL$ is \[\supp(v)=\{\theta_r: E_rv\neq 0\}.\]
For a vertex $a$, we define its support by $\supp(a)=\supp(e_a)$.
We say that two vertices $a$ and $b$ are \textit{strongly cospectral} if, for every $r$, $E_re_a = \pm E_r e_b.$
If $a$ and $b$ are strongly cospectral, we say that their \textit{negative eigenvalue support} is
\[S^-=\{\theta_r \in \supp(a):E_re_a=-E_re_b\}.\]
The \textit{positive eigenvalue support} is defined similarly as $S^+=\{\theta_r \in \supp(a):E_re_a=E_re_b\}.$

Finally, throughout this paper, we use the following characterization of PST proved by Coutinho and Baptista. For convenience, we use an equivalent formulation of their final condition.
\begin{theorem}[{\cite[Theorem~5]{coutinhoQuantumWalksNormalized2024}}]
\label{lemma_characterization_PST}
Let $G$ be a connected graph with vertices $a$ and $b$. Let $\den(r)$ denote the denominator of $r$ when expressed in lowest terms with $\den(r)=1$ if $r$ is an integer. Furthermore, let $S^-$ and $S^+$ respectively denote the negative and positive eigenvalue support of $a$ and $b$. Then, there is perfect state transfer between $a$ and $b$ if and only if the following conditions hold:
\begin{enumerate}
    \item $a$ and $b$ are strongly cospectral.
    \item $S^- \cup S^+ \subseteq \mathbb{Q}$.
    \item For $m = \lcm \{ \den (\theta)\}_{\theta \in S^- \cup S^+}, g = \gcd \{ m \theta\}_{\theta \in S^- \cup S^+}$ and every $\theta \in S^- \cup S^+$,
    \begin{itemize}
        \item $\theta \in S^+$ if and only if $m\theta/g$ is even,
        \item $\theta \in S^-$ if and only if $m\theta/g$ is odd.
    \end{itemize}
\end{enumerate}
 
\noindent If this holds, then the minimum positive time at which there is perfect state transfer between $a$ and $b$ is $\tau = m \pi/g$.
\end{theorem}

\section{PST at odd distance}

In order to show that PST between vertices at odd distance only happens in $K_2$, we use the following result of Coutinho and Baptista, which imposes a  restriction on the negative eigenvalue support of the two vertices. 
\begin{theorem}[{\cite[Theorem~7]{coutinhoQuantumWalksNormalized2024}}]
\label{lemma_different_classes_eigenvalue_rational}
    Let $G$ be a tree on two or more vertices admitting perfect state transfer between vertices $a$ and $b$ at odd distance. Let $S^-$ be their negative eigenvalue support. If $\theta \in S^-$, then $\theta = 2/k$ for some odd positive integer $k$.
\end{theorem}

Furthermore, the following result of Coutinho and Baptista reduces the problem to showing that $2$ is the only eigenvalue in the negative eigenvalue support.
\begin{corollary}[{\cite[Corollary~9]{coutinhoQuantumWalksNormalized2024}}]
    \label{lemma_if_S_only_P2}
    Let $G$ be a tree on two or more vertices admitting perfect state transfer between vertices $a$ and $b$ at odd distance. If $S^-=\{2\}$, then $G = K_2$.
\end{corollary}
Finally, we will need the following small result which is used in~\cite[Corollary~8]{coutinhoQuantumWalksNormalized2024} and whose analogue in the adjacency-matrix case is standard.
\begin{lemma}\label{Lemma_SC_same_degree}
    Let $G$ be a graph with distinct vertices $a$ and $b$. If $a$ and $b$ are strongly cospectral with respect to the normalized Laplacian, then they have the same degree.
 \end{lemma}
 \begin{proof}
     By computation, we can see that $v=D^{1/2} \mathbbm{1}$ where $\mathbbm{1}$ is the vector of all ones, is an eigenvector of $\LL$ for the eigenvalue $0$.
     As $a$ and $b$ are strongly cospectral, we know that $E_0 e_a = \pm E_0 e_b$.
     Hence, $\langle D^{1/2} \mathbbm{1}, e_a\rangle=\pm \langle D^{1/2} \mathbbm{1}, e_b\rangle.$
     However,
     \[\langle D^{1/2} \mathbbm{1}, e_a\rangle = \sqrt{\deg(a)}.\]
     Thus, $\sqrt{\deg(a)}=\pm \sqrt{\deg(b)}$, so $a$ and $b$ have the same degree.
\end{proof}
To show that PST at odd distance is impossible if the tree has more than $2$ vertices, we show that the restriction in Theorem~\ref{lemma_different_classes_eigenvalue_rational} forces the negative eigenvalue support to be $\{2\}$. This establishes the following classification of perfect state transfer at odd distance.

\begin{theorem}\label{theorem_PST_impossible_different_classes}
    The only tree admitting perfect state transfer between vertices at odd distance in the normalized Laplacian model is $K_2$.
\end{theorem}
\begin{proof}
Let $T$ be a tree with two vertices $a$ and $b$ at odd distance that admit perfect state transfer. In particular, $a$ and $b$ are strongly cospectral by Theorem~\ref{lemma_characterization_PST}. We know by Theorem~\ref{lemma_different_classes_eigenvalue_rational} that the elements of $S^-$, the negative eigenvalue support of $a$ and $b$, have the form $\theta=2/k$ with $k$ an odd positive integer.

We first show that $2 \in S^-$. From \cite{coutinhoQuantumWalksNormalized2024}, we know that, in a tree, if $v$ is an eigenvector for $\theta$, then flipping the signs of the entries on one bipartition class gives an eigenvector for $2-\theta$. Since $0$ is an eigenvalue and its eigenspace is spanned by $D^{1/2}\mathbbm{1}$, flipping the signs on one bipartition class gives a vector spanning the eigenspace corresponding to the eigenvalue $2$.

As $a$ and $b$ are at odd distance, they lie in different bipartition classes, so this $2$-eigenvector has entries of opposite signs at $a$ and $b$. Since $a$ and $b$ are strongly cospectral, $E_2 e_a = \pm E_2 e_b$, where $E_2$ denotes the orthogonal projection onto the $2$-eigenspace. Hence
\[E_2e_a=-E_2e_b,\] 
and $2\in S^-$.

To show that $2$ is the only eigenvalue in the negative support of $a$ and $b$, we construct a polynomial $p(x) \in \mathbb{Z}[x]$ such that  $p(\theta)=0$ for every $\theta \in S^-$ and $p(0)=2$.
Then we show that $p$ cannot have roots of the form $2/k$ with odd $k\geq 3$, proving that $S^-=\{2\}$ and thus that $T=K_2$ by Corollary~\ref{lemma_if_S_only_P2}.

Let $L=D-A$ be the combinatorial Laplacian of $T$, so $\LL=D^{-1/2}LD^{-1/2}$, and let $\theta$ be an eigenvalue of $\LL$.
We can see that a vector $v\neq 0$ is an eigenvector for $\theta$ if and only if $(L-\theta D) D^{-1/2}v=0.$ Hence, 
\[\ker(L-\theta D) = \{D^{-1/2}v : v \in \ker(\LL - \theta I)\} \]
Furthermore, if $\theta \in S^-$, then $v_a=-v_b.$
As $a$ and $b$ are strongly cospectral, they have the same degree by Lemma~\ref{Lemma_SC_same_degree}, so $(D^{-1/2}v)_a=-(D^{-1/2}v)_b$. Hence $D^{-1/2}v \perp e_a+e_b$ and thus, for any $\theta \in S^-$,\[\ker(L-\theta D) \perp e_a+e_b.\]

Let $M(x)= L-xD$. By our computations above, we know that for $\theta\in S^-$, $\det(M(\theta))=0.$ Let $\adj(M(x))$ denote the adjugate of the matrix $M(x)$, that is, the transpose of its cofactor matrix. Then, by elementary properties of the determinant, $M(x)\operatorname{adj}(M(x))=\det(M(x))I.$
Therefore, for $\theta\in S^-$,
\[
    M(\theta)\operatorname{adj}(M(\theta))=0,
\]
so every column of $\operatorname{adj}(M(\theta))$ belongs to
$\ker M(\theta)$. Hence, taking $p(x)=(e_a+e_b)^t v(x)$ for any column $v(x)$ of $\adj(M(x))$ implies that $p(\theta)=0$ for $\theta \in S^-$ because then $v(\theta) \perp e_a+e_b$.

We choose the first column for convenience. Thus, let \[p(x)=(e_a+e_b)^t \adj(M(x)) e_1.\]
By definition, $p \in \mathbb{Z}[x]$ as the entries of $\adj(M(x))$ are integer polynomials. Furthermore, the computations above show that
$p(\theta)=0$ for every $\theta \in S^-.$ 

Now, since $T$ is a tree, it has exactly one spanning tree. Hence, by Kirchhoff's Matrix Tree Theorem, $\adj(L)=J$, the matrix of all ones.
Hence, \[p(0)=(e_a+e_b)^t \adj(L) e_1=2.\]
 Thus, as we know that $p(2)=0$, we obtain \[p(x)=(x-2)p_1(x)\] with $p_1(x) \in \mathbb{Z}[x]$ such that $p_1(0)=-1.$

Now, by the Rational Root Theorem, the numerator of the rational roots of $p_1$ must divide $-1$.
Hence, $p_1$ cannot have a root of the form $2/k$ with $k$ odd.

 Thus $S^-=\{2\}$ and by Corollary~\ref{lemma_if_S_only_P2}, $T=K_2$.
\end{proof}
This proof generalizes to give a similar result on bipartite graphs.
\begin{corollary}
    The only bipartite graph with an odd number of spanning trees that admits PST at odd distance is $K_2$.
\end{corollary}
\begin{proof}
    Let $G$ be a bipartite graph with an odd number of spanning trees that admits PST between two vertices $a$ and $b$ at odd distance.
    Reusing the polynomial $p(x)=(e_a+e_b)^t \adj(M(x)) e_1$, the computations remain the same as they were not dependent on the graph being a tree.
    Thus, $p(\theta)=0$ for every $\theta \in S^-$.
    Now, letting $\tau$ be the number of spanning trees, $p(0)=(e_a+e_b)^t \adj(L) e_1= 2 \tau$.
    The sign-switching property giving $2 \in S^-$ is true for any bipartite graphs and so 
    \[p(x)=(x-2)p_1(x)\] where $p_1(x) \in \mathbb{Z}[x]$ such that $p_1(0)=-\tau$ which is odd. Hence, again by the Rational Root Theorem, every eigenvalue in $S^- \setminus \{2\}$ must have odd numerator when in lowest form.
    However, using the characterization in Theorem~\ref{lemma_characterization_PST}, we know that $\theta m/g$ is odd for every $\theta \in S^-.$
    Hence, $h=2m/g$ is odd as $2 \in S^-$. Now, let $\theta \in S^-$. Again, $k=m\theta/g$ is odd. Thus, $\theta=kg/m=2k/h$ which implies that the numerator of any $\theta \in S^-$ is even when in lowest form. This shows that $S^-=\{2\}$.
    
    This implies that
    \[
        e_a-e_b=\sum_r  E_r (e_a-e_b)= E_2 (e_a-e_b)=2E_2 e_a 
        \]
    so $e_a-e_b$ is in the $2$-eigenspace of $\LL$. However, the $2$-eigenspace of a connected bipartite graph is spanned by the vector $D^{1/2}s$, where $s$ is a vector where all coordinates in one bipartition class are $1$ and all the others are $-1$. The coordinates of this vector are all nonzero. Since $e_a-e_b$ is $0$ in every coordinate except $a$ and $b$, the graph can have no other vertices. Hence, the graph has two vertices and is $K_2$.
\end{proof}

\section{Trees with PST at even distance}\label{Sec:examples}

Coutinho and Baptista proved the following necessary condition for perfect state transfer between vertices at even distance in a tree.
\begin{theorem}[{\cite[Corollary~8]{coutinhoQuantumWalksNormalized2024}}]
\label{lemma_same_class_implies_leaves}
    Let $G$ be a tree on three or more vertices, admitting perfect state transfer between vertices $a$ and $b$. If $a$ and $b$ are at even distance, then they are leaves attached to the same vertex. 
\end{theorem}

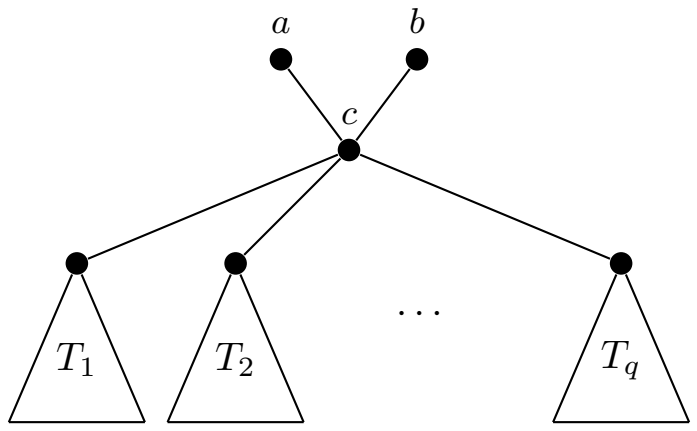
\begin{figure}[ht]
    \begin{center}
        \begin{tikzpicture}[
            scale=0.75,
            transform shape,
            vertex/.style={circle, fill=black, inner sep=2pt},
            every label/.style={font=\small}
        ]

        \node[vertex, label=above:$a$] (a) at (-0.6,2.4) {};
        \node[vertex, label=above:$b$] (b) at (0.6,2.4) {};
        \node[vertex, label=above:$c$] (c) at (0,1.6) {};
        \draw (a) -- (c);
        \draw (b) -- (c);
        
        \node[vertex] (r1) at (-2.4,0.6) {};
        \node[vertex] (r2) at (-1.0,0.6) {};
        \node[vertex] (rq) at (2.4,0.6) {};
        \draw (c) -- (r1);
        \draw (c) -- (r2);
        \draw (c) -- (rq);
        
        \draw (r1) -- (-3.0,-0.8);
        \draw (r1) -- (-1.8,-0.8);
        \draw (-3.0,-0.8) -- (-1.8,-0.8);
        \node at (-2.4,-0.25) {$T_1$};
        
        \draw (r2) -- (-1.6,-0.8);
        \draw (r2) -- (-0.4,-0.8);
        \draw (-1.6,-0.8) -- (-0.4,-0.8);
        \node at (-1.0,-0.25) {$T_2$};
        
        \node at (0.65,0.15) {$\cdots$};
        
        \draw (rq) -- (1.8,-0.8);
        \draw (rq) -- (3.0,-0.8);
        \draw (1.8,-0.8) -- (3.0,-0.8);
        \node at (2.4,-0.25) {$T_q$};
    \end{tikzpicture}
    \end{center}
        \caption{All trees with more than $2$ vertices admitting PST have this form.}
        \label{Figure_generic_tree}
\end{figure}

Combining this result with Theorem~\ref{theorem_PST_impossible_different_classes}, we know that if a tree on more than $2$ vertices admits PST between two vertices $a$ and $b$, it has the form shown in Figure~\ref{Figure_generic_tree} where $q\geq 0$ and $T_1,...,T_q$ are rooted trees.

We construct an infinite family of trees of this form admitting PST between $a$ and $b$, proving that this condition can indeed be realized by a non-trivial tree and that PST can happen in arbitrarily large trees.

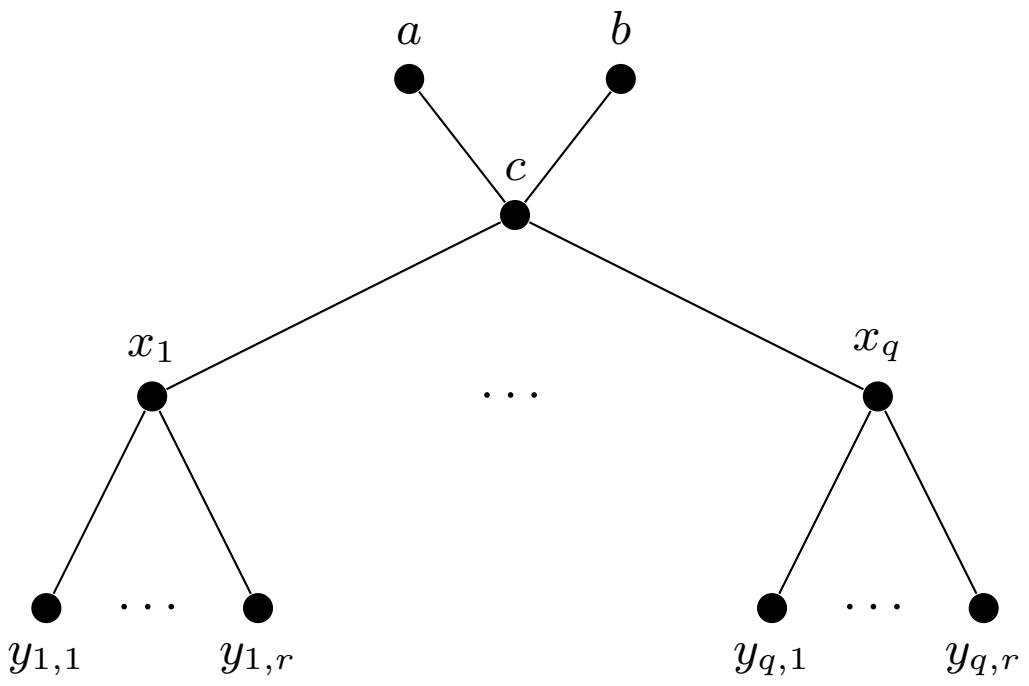
\begin{figure}[ht]
    \begin{center}
        \begin{tikzpicture}[
            vertex/.style={circle, fill=black, inner sep=2pt},
            every label/.style={font=\small}
        ]
        
        \node[vertex, label=above:$a$] (a) at (-0.7,3.2) {};
        \node[vertex, label=above:$b$] (b) at (0.7,3.2) {};
        \node[vertex, label=above:$c$] (c) at (0,2.3) {};
        \draw (a) -- (c);
        \draw (b) -- (c);
        
        \node[vertex, label=above:$x_1$] (x1) at (-2.4,1.1) {};
        \node at (0,1.1) {$\cdots$};
        \node[vertex, label=above:$x_q$] (xq) at (2.4,1.1) {};
        \draw (c) -- (x1);
        \draw (c) -- (xq);
        
        \node[vertex, label=below:$y_{1,1}$] (y11) at (-3.1,-0.3) {};
        \node at (-2.4,-0.3) {$\cdots$};
        \node[vertex, label=below:$y_{1,r}$] (y1r) at (-1.7,-0.3) {};
        \draw (x1) -- (y11);
        \draw (x1) -- (y1r);
        
        \node[vertex, label=below:$y_{q,1}$] (yq1) at (1.7,-0.3) {};
        \node at (2.4,-0.3) {$\cdots$};
        \node[vertex, label=below:$y_{q,r}$] (yqr) at (3.1,-0.3) {};
        \draw (xq) -- (yq1);
        \draw (xq) -- (yqr);
        \end{tikzpicture}
    \end{center}
    \caption{For certain values of $q$ and $r$, this tree admits PST between $a$ and $b$.}\label{Figure_Tree_Tqr}
\end{figure}

For positive integers $q$ and $r$ we define $T(q,r)$ to be the tree on $q(r+1)+3$ vertices shown in Figure~\ref{Figure_Tree_Tqr}.
Formally, $T(q,r)=(V,E)$ where
\begin{itemize}
    \item $V=\{a,b,c\} \cup \{x_i,y_{i,j} :1 \leq i \leq q, 1 \leq j \leq r \}$, and
    \item $E=\{\{a,c\},\{b,c\}\} \cup \{\{c,x_i\},\{x_i,y_{i,j}\}: 1 \leq i \leq q, 1\leq j \leq r\}$.
\end{itemize}

We show that, under certain number theoretic conditions on $q$ and $r$ that are satisfied for arbitrarily large values, $T(q,r)$ admits perfect state transfer between $a$ and $b$. 
Our first result gives one characterization of the pairs $q, r$ for which $T(q,r)$ admits PST between $a$ and $b$. After this theorem, we give the characterization in a more useful way.
\begin{theorem}\label{Theorem_T_p_has_PST} For positive integers $q$ and $r$, the tree 
    $T(q,r)$ admits perfect state transfer between $a$ and $b$ if and only if, in lowest form, \[\frac{2r}{(q+2)(r+1)}=\frac{u^2}{v^2},\] for positive odd integers $u$ and $v$.
    Furthermore, if there is PST, the minimum positive time at which it occurs is $v\pi$.
\end{theorem}
\begin{proof}

In order to use Theorem~\ref{lemma_characterization_PST}, we start by describing the eigenvalue support of vertices $a$ and~$b$. As these vertices are indistinguishable in $T(q,r)$, we examine $e_a-e_b$ and $e_a+e_b$ to get the information we need.
Hence, let
\[
    \alpha=e_a-e_b,
    \qquad
    \beta=e_a+e_b,
\]
and let $R=D^{-1/2}AD^{-1/2},$ so that $\LL=I-R$.

\smallskip
\noindent \textit{Eigenvalue support of $\alpha$}.  
    As $a$ and $b$ are both leaves and have a common neighbor, we have $R\alpha=0$ and thus $\LL \alpha =(I-R)\alpha = \alpha.$
    Therefore, $\alpha$ is a $1$-eigenvector of $\LL$ and $\supp(\alpha)=\{1\}.$

\smallskip
\noindent \textit{Eigenvalue support of $\beta$}. Let \[x=\sum_{i=1}^q e_{x_i}, \quad y=\sum_{i=1}^q\sum_{j=1}^r e_{y_{i,j}}.\]
    We show that $W= \spn(\beta, e_c , x,y) $ is $\LL$-invariant. Then, computing the eigenvalues of the restriction of $\LL$ to $W$ gives the support of $\beta$. Direct computation gives
    \begin{align*}
        &R \beta  = \frac{2}{\sqrt{q+2}} \, e_c,  &&R e_c = \frac{1}{\sqrt{q+2}}\, \beta + \frac{1}{\sqrt{(q+2)(r+1)}}\, x,\\ &Rx= \frac{q}{\sqrt{(q+2)(r+1)}}\, e_c+\frac{1}{\sqrt{r+1}}\, y, &&Ry= \frac{r}{\sqrt{r+1}}\,  x.
    \end{align*}

Thus, $W$ is $R$-invariant  and thus $\LL$-invariant.
Furthermore, we see that, in the orthogonal basis $(\beta, e_c, x, y)$, the restriction of $R$ to $W$ is
\begin{align*}
        R|_W =
\begin{pmatrix}
    0 & \frac{1}{\sqrt{q+2}} & 0 & 0\\
    \frac{2}{\sqrt{q+2}} & 0 & \frac{q}{\sqrt{(q+2)(r+1)}} & 0\\
    0 & \frac{1}{\sqrt{(q+2)(r+1)}} & 0 & \frac{r}{\sqrt{r+1}}\\
    0 & 0 & \frac{1}{\sqrt{r+1}} & 0
\end{pmatrix}.
    \end{align*}
Letting \[s=\sqrt{\frac{2r}{(q+2)(r+1)}},\] the eigenvalues of $R|_W$ are 
$-1, -s, s$ and $ 1.$
As $\LL=I-R$, this means that the eigenvalues of $\LL|_W$ are $0,1-s,1+s$ and $2$.
Hence, $\supp(\beta) \subseteq \{0,1-s,1+s,2\}.$

Now, let $v\neq 0$ be an eigenvector for some $\theta \in \{0,1-s,1+s,2\}$ written in the basis used earlier. Assume for a contradiction that $v_1=0$ and thus that $\beta$ has $0$ projection onto $v$.
As $\LL|_W v =  \theta v$ and $v_1=0$, we get that $(\LL|_W v)_1=-1/\sqrt{q+2} v_2 =0$. Hence, $v_2=0$.
This in turn implies that $(\LL|_W v)_2=-q/\sqrt{(q+2)(r+1)} v_3 =0$ and thus that $v_3=0$.
Finally, this implies that $(\LL|_W v)_3 = -r/\sqrt{r+1} v_4=0$.
Therefore, $v=0$ which is a contradiction.
This proves that $v_1 \neq 0$ and thus that $\theta$ is in the support of $\beta$. As $\theta$ was chosen arbitrarily, we get that $\supp(\beta)=\{0,1-s,1+s,2\}.$

We now have sufficient information to determine when PST occurs between $a$ and $b$. As $e_a= (\alpha + \beta)/2 $ and $\alpha$ and $\beta$ have disjoint supports, we get that $\supp(a)$ is the disjoint union of $\supp(\alpha)$ and $\supp(\beta)$.
    Thus \begin{align*}
    \supp(a) = \left\{0, 1-s,1, 1+s,2\right\}.
\end{align*} 
Furthermore, as $e_b = 
(\beta - \alpha)/2$, $\supp(a)=\supp(b)$.

We now determine when the conditions in Theorem~\ref{lemma_characterization_PST} hold.

\begin{enumerate}
    \item Let $\theta_r \in \supp(a)$ then, either $\theta_r \in \supp(\alpha)$ and
    \[E_r e_a = E_r \frac{\alpha+\beta}{2}= -E_r \frac{\beta-\alpha}{2}=-E_r e_b,\]
or $\theta_r \in \supp(\beta)$ and we then have
\[E_r e_a = E_r \frac{\alpha+\beta}{2}= E_r \frac{\beta-\alpha}{2}=E_r e_b.\]
    Thus, $a$ and $b$ are strongly cospectral. Furthermore, $S^-=\{1\}$ and $S^+=\{0, 1-s, 1+s,2\}$.
    \item As $\supp(a)= \{0, 1-s,1, 1+s,2\}$, $\supp(a)\subseteq \mathbb{Q}$ if and only if $s$ is rational. This is true if and only if $s^2$ has the form $u^2/v^2$ for some $u$ and $v$ when in lowest form.
    \item Assuming that the previous condition is satisfied and thus that $s=u/v$, as $\supp(a)= \{0,1-s,1,1+s,2\}$, we have
    $m=v$ and $g=1.$
    Thus, the last condition of the theorem is satisfied if and only if 
    \begin{align*}
        &v\cdot \theta \text{ is odd for } \theta \in \{1\}, \text{ and}\\
        &v \cdot \theta \text{ is even for } \theta \in \{0,1-s,1+s,2\}.
    \end{align*}
    Therefore, for this condition to be satisfied, $v$ has to be odd.
    The condition is automatically true for $\theta= 0$ and $2$.
    Now, when $\theta=1\pm s$, 
    $m\theta=v(1\pm u/v)=v \pm u$. This number is even if and only if $u$ has the same parity as $v$.
Therefore, the third condition of the theorem is satisfied if and only if \[s^2=\frac{2r}{(q+2)(r+1)}\] has the form $u^2/v^2$ with $u$ and $v$ both odd. 
\end{enumerate}

We conclude that there is PST between $a$ and $b$ if and only if \[2r/((q+2)(r+1)) = u^2/v^2\] with $u$ and $v$ both odd when in lowest form. Furthermore, if PST does occur, then the minimum positive time at which it happens is $v\pi$.
\end{proof}

While Theorem~\ref{Theorem_T_p_has_PST} gives a characterization of the pairs $q,r$ for which $T(q,r)$ admits PST, it is not clear that the condition is satisfied by arbitrarily large trees. The following result restates the condition on $q$ and $r$ in a form which directly shows that the trees can be arbitrarily large. This new characterization is also more useful as it makes it easier to find pairs $q,r$ for which $T(q,r)$ admits PST by fixing the value of $r$ and exhibiting all the values of $q$ which satisfy the condition.
\begin{theorem}
    Let $r\geq 1$, and let $d$ be the product of the odd primes occurring to an odd exponent in the prime factorization of $r(r+1)$. Then we have the following.
    \begin{enumerate}
        \item If $r$ is even of the form $r=2^t r_0$
        where $t\geq 1$ and $r_0$ is odd, then $T(q,r)$ admits PST between $a$ and $b$ if and only if
        \[
            q=2^{t+1}d n^2-2
        \]
        for some odd $n\geq 1$.
    
        \item If $r\equiv 1\pmod{4}$, then $T(q,r)$ admits PST between $a$ and $b$ if and only if
        \[
        q=d n^2-2\geq 1
        \]
        for some odd $n\geq 1$.
    
        \item If $r\equiv 3\pmod{4}$, then $T(q,r)$ does not admit PST between $a$ and $b$ for any choice of $q$.
    \end{enumerate}
\end{theorem}
\begin{proof}
    Let $\tilde q=q+2$. We know by Theorem~\ref{Theorem_T_p_has_PST} that there is PST between $a$ and $b$ in $T(q,r)$ if and only if 
    \[
    \frac{2r}{\tilde q(r+1)}=\frac{u^2}{v^2},
    \]
    for some odd $u$ and $v$ when in lowest form.

    \textit{Case 1:} If $r$ is even and of the form $r=2^t r_0$, the fraction on the left is equal to 
    \[\frac{2^{t+1}r_0}{\tilde q(r+1)}.\]
    Let $d_1$,$ d_2$ be the products of the primes occurring to an odd exponent in $r_0$ and $r+1$ respectively. Thus, $r=2^t r_0=2^t d_1 h^2$, $r+1=d_2  m^2$ for some odd $h$ and $m$. 
    Hence the fraction above is equal to 
    \[\frac{2^{t+1}d_1 h^2}{\tilde q d_2 m^2}\]
As $d_1 h^2$ and $d_2m^2$ are coprime because $r$ and $r+1$ are, this implies that it is necessary and sufficient for $\tilde{q}$ to have the form $2^{t+1}d_1d_2 n^2$ for some odd $n\geq 1$ for the fraction to have the desired form when in lowest terms. By definition of $d_1$ and $d_2$, $d_1d_2=d$, the product of the odd primes occurring to an odd exponent in the prime factorization of $r(r+1)$.
Hence, if $r$ is even, there is PST in $T(q,r)$ if and only if $\tilde q=q+2$ has the form $2^{t+1}dn^2$ with $n\geq 1$ odd.

    \textit{Case 2:} If $r$ is odd, then $r+1$ is even and we have
    \[\frac{2r}{\tilde q(r+1)}=\frac{r}{\tilde q \frac{r+1}{2}}.\]
    As $r$ is odd, for this fraction to be the quotient of two odd squares, it is necessary for $(r+1)/2$ to be odd and thus $r \equiv 1 \pmod{4}$.
    Let $d_1,d_2$ be the products of the odd primes occurring to an odd exponent in $r$ and $r+1$ respectively.
    Hence, $r=d_1 h^2$ and $r+1=2d_2m^2$ for some odd $h$ and $m$.
    The fraction above is thus equal to 
    \[\frac{d_1 h^2}{\tilde{q} d_2 m^2}.\]
    Again, as $r$ and $r+1$ are coprime, $d_1 h^2$ and $d_2m^2$ are as well.
    Thus it is necessary and sufficient for $\tilde{q}$ to have the form $d_1d_2n^2$ for some odd $n\geq 1$, for the fraction to have the desired form when in lowest form.
    Again, $d_1d_2=d$ and  we can conclude that if $r \equiv 1 \pmod{4}$, then $T(q,r)$ has PST between $a$ and $b$ if and only if $q=dn^2-2$ for some odd $n \geq 1$.
    Furthermore, if $r \equiv 3 \pmod{4}$ then no choice of $q$ yields PST in $T(q,r)$.

\end{proof}

\section{Further examples}\label{Sec:more_examples}

The family in the previous section shows that PST can happen in arbitrarily large trees. Coutinho and Baptista proved a characterization of this phenomenon with the spectral condition in Theorem~\ref{lemma_characterization_PST}, but it would be nice to have a structural characterization. However, the examples of this section seem to indicate that obtaining such a characterization may be difficult.

\begin{figure}[ht]
\centering

\resizebox{0.65\textwidth}{!}{%
\begin{tikzpicture}[
    vertex/.style={
        circle,
        fill=black,
        inner sep=2.2pt
    },
    labelnode/.style={
        fill=none,
        draw=none,
        inner sep=1pt
    }
]


\node[vertex] (c) at (0,0) {};
\node[labelnode, above=3pt of c] {$c$};

\node[vertex] (a) at (-1.5,1.4) {};
\node[vertex] (b) at (1.5,1.4) {};

\draw (c)--(a);
\draw (c)--(b);

\node[labelnode, above left=2pt of a] {$a$};
\node[labelnode, above right=2pt of b] {$b$};


\node[vertex] (x1) at (-3.5,-2.1) {};
\node[vertex] (xlast) at (3.5,-2.1) {};

\draw (c)--(x1);
\draw (c)--(xlast);

\node[labelnode, above=3pt of x1] {$x_1$};
\node[labelnode, above=3pt of xlast] {$x_{q^2-2}$};

\node[labelnode] at (0,-1.45) {$\cdots$};


\node[vertex] (u11) at (-4.55,-1.15) {};
\node[vertex] (u12) at (-4.75,-2.05) {};
\node[vertex] (u13) at (-4.55,-2.95) {};

\draw (x1)--(u11);
\draw (x1)--(u12);
\draw (x1)--(u13);

\node[labelnode, above left=2pt of u11] {$u_{1,1}$};
\node[labelnode, left=6pt of u12] {$u_{1,2}$};
\node[labelnode, below left=2pt of u13] {$u_{1,3}$};


\node[vertex] (ur1) at (4.55,-1.15) {};
\node[vertex] (ur2) at (4.75,-2.05) {};
\node[vertex] (ur3) at (4.55,-2.95) {};

\draw (xlast)--(ur1);
\draw (xlast)--(ur2);
\draw (xlast)--(ur3);

\node[labelnode, above right=2pt of ur1]
    {$u_{q^2-2,1}$};

\node[labelnode, right=6pt of ur2]
    {$u_{q^2-2,2}$};

\node[labelnode, below right=2pt of ur3]
    {$u_{q^2-2,3}$};


\node[vertex] (y11) at (-4.6,-4.15) {};
\node[vertex] (y1last) at (-2.4,-4.15) {};

\draw (x1)--(y11);
\draw (x1)--(y1last);

\node[labelnode, left=5pt of y11] {$y_{1,1}$};
\node[labelnode, right=5pt of y1last]
    {$y_{1,q^2-2}$};

\node[labelnode] at (-3.5,-4.0) {$\cdots$};


\node[vertex] (yr1) at (2.4,-4.15) {};
\node[vertex] (yrlast) at (4.6,-4.15) {};

\draw (xlast)--(yr1);
\draw (xlast)--(yrlast);

\node[labelnode, left=5pt of yr1]
    {$y_{q^2-2,1}$};

\node[labelnode, right=5pt of yrlast]
    {$y_{q^2-2,q^2-2}$};

\node[labelnode] at (3.5,-4.0) {$\cdots$};


\node[vertex] (z11) at (-4.6,-5.9) {};
\node[vertex] (z1last) at (-2.4,-5.9) {};

\draw (y11)--(z11);
\draw (y1last)--(z1last);

\node[labelnode, below left=3pt of z11]
    {$z_{1,1}$};

\node[labelnode, below right=3pt of z1last]
    {$z_{1,q^2-2}$};


\node[vertex] (zr1) at (2.4,-5.9) {};
\node[vertex] (zrlast) at (4.6,-5.9) {};

\draw (yr1)--(zr1);
\draw (yrlast)--(zrlast);

\node[labelnode, below left=3pt of zr1]
    {$z_{q^2-2,1}$};

\node[labelnode, below right=3pt of zrlast]
    {$z_{q^2-2,q^2-2}$};

\end{tikzpicture}%
}

\caption{The tree $U(q)$, which admits PST between $a$ and $b$ under certain conditions on $q$.}
\label{Figure_Tree_U}
\end{figure}
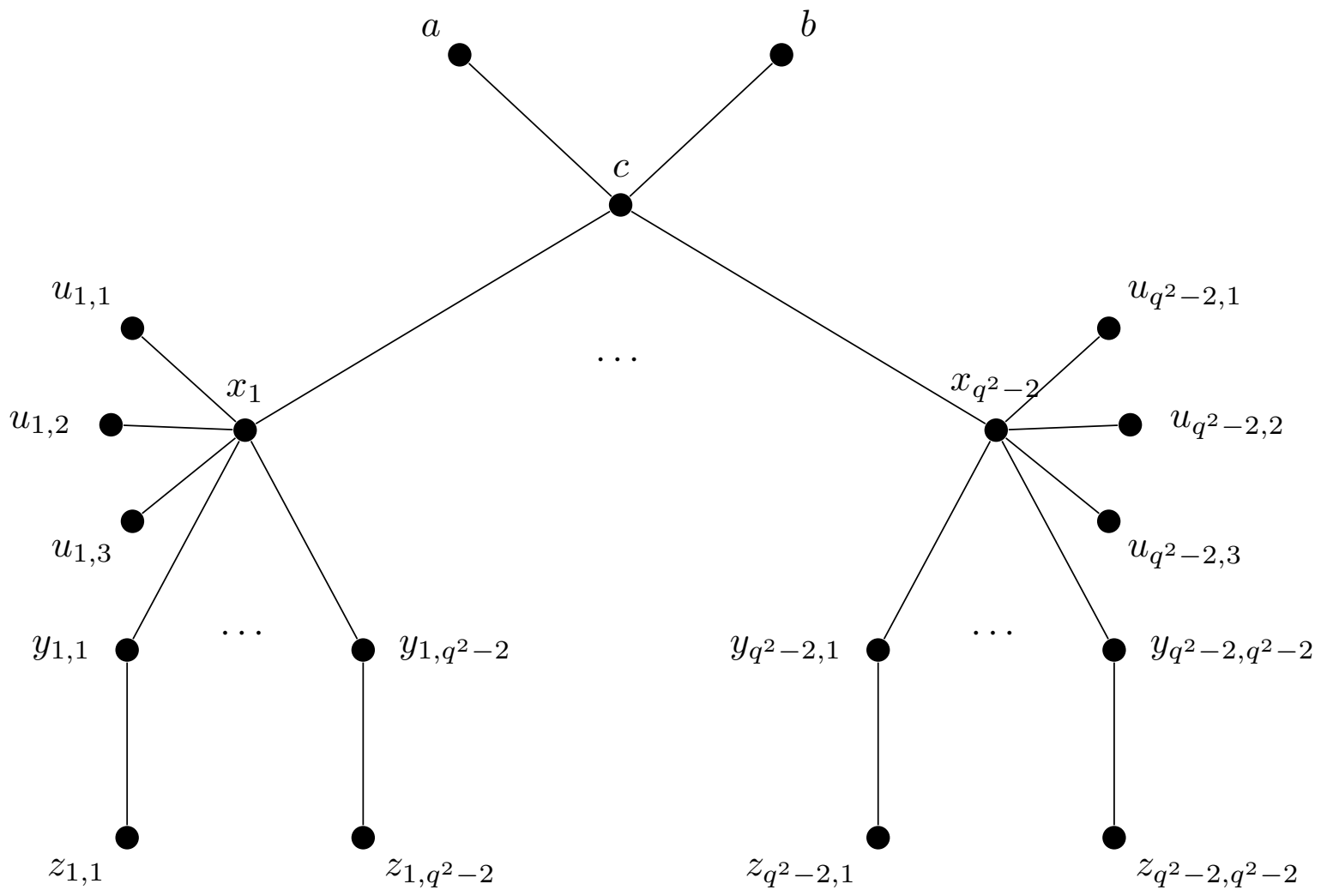

For our next example of trees with PST, let $q\geq 3$ be an integer and define $U(q)$ to be the tree on $3+2q^2(q^2-2)$ vertices shown in Figure~\ref{Figure_Tree_U}.

\begin{theorem}\label{Theorem_U_has_PST} For $q\geq 3$,
    $U(q)$ admits PST between $a$ and $b$ if and only if $(q^2+2)/3$ is a perfect square.
    Furthermore, if PST does occur, then the minimum positive time at which it occurs is \[\sqrt{\frac{q^2+2}{3}}\  q\pi.\]
\end{theorem}
\begin{proof}
    The proof is very similar to the proof of Theorem~\ref{Theorem_T_p_has_PST} except that in this case, $\LL|W$ is not tridiagonal so a slightly different argument is needed to show that its eigenvalues all lie in the support of $\beta$.
    Let $R=D^{-1/2}AD^{-1/2}$ such that $\LL = I - R$.
    Let $\alpha=e_a-e_b$ and $\beta= e_a+e_b$.
    Again, $\alpha$ is a $1$-eigenvector of $\LL$, so $\supp(\alpha)=\{1\}.$

    Now, let \[x=\sum_{i=1}^{q^2-2} e_{x_i},\quad u=\sum_{i=1}^{q^2-2} \left( e_{u_{i,1}}+e_{u_{i,2}}+e_{u_{i,3}}\right),\quad y = \sum_{i=1}^{q^2-2}\sum_{j=1}^{q^2-2} e_{y_{i,j}},\quad z=\sum_{i=1}^{q^2-2}\sum_{j=1}^{q^2-2} e_{z_{i,j}}.\]
    By computation, we can check that $W=\spn(  \beta, e_c, x, u, y, z)$ is $R$-invariant and thus $\LL$-invariant.
    Furthermore, in this orthogonal basis, 
    \[
\LL|_W
=
\begin{pmatrix}
1 & -\frac1q & 0 & 0 & 0 & 0\\[2mm]
-\frac2q & 1 &
-\frac{q^2-2}{q\sqrt{q^2+2}} & 0 & 0 & 0\\[2mm]
0 &-\frac{1}{q\sqrt{q^2+2}}&1&
-\frac{3}{\sqrt{q^2+2}}&
-\frac{q^2-2}{\sqrt{2(q^2+2)}}&0\\[2mm]
0&0&-\frac1{\sqrt{q^2+2}}&1&0&0\\[2mm]
0&0&-\frac1{\sqrt{2(q^2+2)}}&0&1&-\frac1{\sqrt2}\\[2mm]
0&0&0&0&-\frac1{\sqrt2}&1
\end{pmatrix}.
\]
Its eigenvalues are $0,1\pm 1/q,  1\pm\sqrt{3/(q^2+2)}$ and $2$.

We prove that they are all in the support of $\beta$. Let $v\neq 0$ be an eigenvector for $\theta \in \{0,1\pm 1/q,  1\pm\sqrt{3/(q^2+2)},2\}$ written in the same basis as above. Suppose for a contradiction that $v_1=0$.
A similar argument as the one in Theorem~\ref{Theorem_T_p_has_PST} gives $v_1=v_2=v_3=0$.
Then, $(\LL|_W v)_4= v_4 = \theta v_4 $. As $\theta\neq 1$ we obtain that $v_4=0$.
Therefore, $(\LL v)_3= -(q^2-2)/\sqrt{2(q^2+2)} v_5 = 0$ and so $v_5=0$.
Finally, $(\LL v)_6 = \theta v_6 = v_6$, so, as $\theta \neq 1$, $v_6=0$.
This implies that $v=0$ which is a contradiction.
Hence, $v_1\neq 0$ and $\theta$ is in the support of $\beta$.
Thus, $\supp(\beta)=\{0,1\pm 1/q,  1\pm\sqrt{3/(q^2+2)},2\}$.

Thus, the support of $a$ is $\{0,1, 1\pm 1/q,  1\pm\sqrt{3/(q^2+2)},2\}$ and the negative eigenvalue support of $a$ and $b$ is $\{1\}$ while the positive support is $\{0,1\pm 1/q,  1\pm\sqrt{3/(q^2+2)},2\}.$

Theorem~\ref{lemma_characterization_PST} then implies that to get PST between $a$ and $b$, $\sqrt{3/(q^2+2)}$ has to be rational, so 
\[\frac{3}{q^2+2}=\frac{N^2}{M^2}\]
for some $N,M  \geq 1$ when in lowest form. Hence, $N^2 \mid 3$ which implies that $N=1$.
So, to get PST, we need $\sqrt{(q^2+2)/3}$ to be an integer and thus $(q^2+2)/3$ to be the square of an integer.

Finally, if the previous condition holds, such that 
\[\sqrt{\frac{q^2+2}{3}}=h\]for some positive integer $h$, then $m=\lcm(q,h)$. As $q^2+2=3h^2$, any common divisor of $q$ and $h$ must also divide $2$. Hence, $\gcd(q,h)=1$ or $2$. However, if $q$ was even, then $q^2+2 \equiv 2\mod 4$ while $3h^2 \equiv 0 \text{ or } 3 \mod 4$. Thus, $q$, and thus also $h$, are odd and $\gcd(q,h)=1$. Therefore, 
\[m=qh \quad \text{and} \quad g=1.\]
The last condition of Theorem~\ref{lemma_characterization_PST} therefore implies that there is PST between $a$ and $b$ if and only if $q$ and $h=\sqrt{(q^2+2)/3}$ are odd. It was established earlier that these conditions are automatically satisfied when $\sqrt{(q^2+2)/3}$ is an integer.

We conclude that there is PST between $a$ and $b$ if and only if $(q^2+2)/3$ is a perfect square.
\end{proof}

The condition on $q$ is satisfied by arbitrarily large values of $q$, which implies that this family is indeed infinite.
To see this, we can see that to have $(q^2+2)/3$ be a perfect square, we need $(q^2+2)/3=h^2$ for some $h$, and thus that $q^2-3h^2=-2$.
This equation is true for the pair $(h,q)=(1,1)$. Furthermore, if it is satisfied by a pair $(h,q)$ then it is also satisfied by $(2h+q,3h+2q)$.
This recurrence produces increasing $q$'s which verify the property.

Finally, we construct a large tree admitting PST between $a$ and $b$ which does not belong to either of the infinite families constructed above. This tree has the property that the rooted trees connected to $c$ have height $3$, making it plausible that arbitrary height can be achieved.

Let $T$ be the tree on $17,164,094,977$ vertices defined as follows. The vertex $c$ is adjacent to the two leaves $a$ and $b$ and to $12,127$ other vertices
\[
    x_1,\ldots,x_{12127}.
\]
For every $i$, $x_i$ is adjacent to $311$ other vertices
\[
    y_{i,1},\ldots,y_{i,311}.
\]
For every $i$ and $j$, $y_{i,j}$ is adjacent to $350$ other vertices
\[
    z_{i,j,1},\ldots,z_{i,j,350}.
\]
Finally, for each $i,j$ and $k$, the vertex $z_{i,j,k}$ is adjacent to $12$ leaves
\[
    w_{i,j,k,1},\ldots,w_{i,j,k,12}.
\]

\begin{theorem}
    $T$ admits PST between $a$ and $b$ and the minimum positive time at which it occurs is $117 \pi$.
\end{theorem}
\begin{proof}
    The proof is similar to the proofs of Theorems~\ref{Theorem_T_p_has_PST} and~\ref{Theorem_U_has_PST}.
    Let $R=D^{-1/2}AD^{-1/2}$ so that $\LL = I - R$.
    Let $\alpha=e_a-e_b$ and $\beta= e_a+e_b$.
    Again, $\alpha=e_a-e_b$ is a $1$-eigenvector for $\LL$ and thus $\supp(\alpha)=\{1\}$.

    Let $x$, $y$, $z$, and $w$ denote the sums of the standard basis vectors corresponding to the vertices $x_i$, $y_{i,j}$, $z_{i,j,k}$, and $w_{i,j,k,l}$, respectively.
    We can check computationally that $W=\spn( \beta, e_c, x, y, z, w)$ is $R$-invariant and that, in the orthonormal basis $( \beta/\|\beta\|, e_c, x/\|x\|, y/\|y\|, z/\|z\|, w/\|w\|)$,
    \[
R|_W=
\begin{pmatrix}
0 & \sqrt{\frac2{12129}} &0&0&0&0\\
\sqrt{\frac2{12129}}&0&
\sqrt{\frac{12127}{12129\cdot312}}&0&0&0\\
0&\sqrt{\frac{12127}{12129\cdot312}}&0&
\sqrt{\frac{311}{312\cdot351}}&0&0\\
0&0&\sqrt{\frac{311}{312\cdot351}}&0&
\sqrt{\frac{350}{351\cdot13}}&0\\
0&0&0&\sqrt{\frac{350}{351\cdot13}}&0&
\sqrt{\frac{12}{13}}\\
0&0&0&0&\sqrt{\frac{12}{13}}&0
\end{pmatrix}.
\]
Hence, $R|_W$ has eigenvalues $\pm 1, \pm 1/13, \pm 1/117$ and $\LL|_W$ has eigenvalues $0,1-1/13, 1-1/117, 1+1/117,1+1/13$ and $2$.

Again, using the same argument as in Theorem~\ref{Theorem_T_p_has_PST}, one can check that all these eigenvalues are in the support of $\beta$. We now have all the needed information on the positive and negative eigenvalue support of $a$ and $b$ to determine that Theorem~\ref{lemma_characterization_PST} holds, with $m=117$ and $g=1$. Therefore, there is PST between $a$ and $b$ and the minimum positive time at which it occurs is $117 \pi$.
\end{proof}

\section{Concluding remarks}

Combining our Theorem~\ref{theorem_PST_impossible_different_classes} and Coutinho and Baptista's Theorem~\ref{lemma_same_class_implies_leaves}, we obtain the following result.

\begin{corollary}\label{Theorem_Classification_pairs}
    Let $T$ be a tree with two or more vertices admitting PST between vertices $a$ and $b$. Then either 
    \begin{itemize}
        \item $a$ and $b$ are both leaves attached to the same vertex, or
        \item $T$ is $K_2$.
    \end{itemize}
\end{corollary}
Moreover, the examples of Section~\ref{Sec:examples} show that the first case of Corollary~\ref{Theorem_Classification_pairs} occurs for arbitrarily large trees, while the examples of Section~\ref{Sec:more_examples} illustrate the difficulty of obtaining a characterization of such trees.

\end{document}